\documentclass[aps,prl,twocolumn,superscriptaddress,amsmath,amssymb,preprintnumbers,showpacs]{revtex4}

\usepackage{graphicx}
\usepackage{dcolumn}

\begin{document}


\title{Limits in the characteristic function description of non-Lindblad-type open quantum systems}


\author{Sabrina Maniscalco}

\affiliation{School of Pure and Applied Physics, University of
KwaZulu-Natal, Durban 4041, South Africa}
\email{maniscalco@ukzn.ac.za}

\affiliation{INFM, MIUR and Dipartimento di Scienze Fisiche ed
Astronomiche dell'Universit\`{a} di Palermo, via Archirafi 36,
90123 Palermo, Italy}

\date{\today}

\begin{abstract}
In this paper I investigate the usability of the characteristic
functions for the description of the dynamics of open quantum
systems focussing on non-Lindblad-type master equations. I
consider, as an example, a non-Markovian generalized master
equation containing a memory kernel which may lead to nonphysical
time evolutions characterized by negative values of the density
matrix diagonal elements [S.M. Barnett and S. Stenholm, Phys. Rev.
A {\bf 64}, 033808 (2001)]. The main result of the paper is to
demonstrate that there exist situations in which the symmetrically
ordered characteristic function is perfectly well defined while
the corresponding density matrix loses positivity. Therefore
nonphysical situations may not show up in the characteristic
function. As a consequence, the characteristic function cannot be
considered an {\it alternative complete} description of the
non-Lindblad dynamics.
\end{abstract}


\maketitle

The theory of open quantum systems describes the interaction of a
quantum system with its environment \cite{petruccionebook}.
Although many physical systems may be considered under certain
conditions quasi-closed for certain intervals of time, no quantum
system may be seen as completely isolated from its surroundings.
The unavoidable interaction between the system and its environment
leads to the phenomena of decoherence and dissipation
\cite{giulini}.

The description of the dynamics of open systems has recently
attracted much of attention for mainly two reasons. On the one
hand environment induced decoherence due to the establishment of
entanglement between the system and the environment is one of the
key issues of the quantum measurement theory \cite{giulini}. On
the other hand, the system-environment interaction seems to be the
major limiting factor in the realization of quantum devices
necessary for the new quantum technologies, e.g. quantum
computation \cite{nielsen}.

The study of the dynamics of an open system is, in general, a very
difficult task. Usually, even if one is interested in the dynamics
of the system only, the influence of the typically infinite
environmental degrees of freedom, makes it impossible to solve
exactly the equations of motion for the observables of interest.
For this reason, the standard description of open systems rely on
a number of approximations which allow to derive a master equation
for the reduced density matrix of the system. The two most common
approximations are the weak coupling approximation, valid when the
interaction between system and environment is sufficiently weak,
and the Markov approximation, relying on the assumption that the
characteristic times of the system are much larger than those of
the environment \cite{alicki}.

Generally, when these approximations are satisfied, the master
equation for the reduced density matrix may be written in the
so-called Lindblad form, which is the only possible form of
first-order linear differential equation, for a completely
positive dynamical semigroup having bounded generator
\cite{Lindblad,Gorini}. The Lindblad master equation, however, is
valid as long as the weak coupling and Markov approximation hold.
While these assumptions are often well justified in quantum
optics, in many solid-state systems, i.e. photonic band-gap
materials and quantum dots, the Markov approximation does not hold
\cite{quang97}. Similarly, the reservoir interacting with a single
mode cavity in atom lasers is strongly non-Markovian
\cite{hope00}. Non-Markovian generalized master equations usually
ar not of Lindblad type.


It is worth noticing that there exist also Markovian systems
described by master equations which cannot be cast in the Lindblad
form \cite{Zurek89,Munro96}. An important problem in the
description of open quantum systems whose master equations are not
in the Lindblad form is that their dynamical map needs not be
completely positive \cite{kraus}, and this may lead to physical
inconsistency. It may even happen that the positivity condition of
the density matrix during the time evolution, a condition less
restrictive then complete positivity but necessary to guarantee
the probabilistic interpretation of the density matrix, breaks
down.

For the sake of completeness, let me underline that complete
positivity is a necessary requirement for a consistent physical
description of open quantum systems whenever factorized initial
conditions for the system and the reservoir are assumed, i.e.
$\hat{\rho}_{\rm T}(0)= \hat{\rho}(0) \otimes \hat{\rho}_{\rm
E}(0)$, with $\hat{\rho}_{\rm T}(0), \hat{\rho}(0),
\hat{\rho}_{\rm E}(0)$, initial density matrices of the total
system, of the reduced system of interest and of the environment,
respectively. Most of the derivations of master equations found in
the literature rely on this assumption. However, when correlations
are present at the initial time, acceptable quantum dynamics which
are not completely positive may exist \cite{Pechukas94}. In the
following I will focus on the case of factorized initial
conditions for which the dynamical map must be completely
positive.

When working with non-Markovian generalized master equations, or
with master equations which are not in the Lindblad form, it is of
crucial importance to establish conditions under which the density
matrix preserves positivity and complete positivity during the
time evolution. In most of the cases these conditions are given in
terms of the density matrix elements at time $t$, and therefore
require the knowledge of the analytic solution of the
non-Markovian master equation. In Ref. \cite{budini}, conditions
for complete positivity in  terms of the memory kernel are
presented for a class of generalized master equations. To the best
of the author's knowledge, however, a criterion analogous to the
Lindblad one for general master equations with memory has not yet
been formulated.

Very useful tools for the description of the dynamics of
paradigmatic open quantum systems such as the damped harmonic
oscillator (quantized mode of the electromagnetic field, motion of
a trapped ion) or the quantum Brownian particle, are the
characteristic functions and the quasi-probability distribution
functions. Both of them contain all the information necessary to
reconstruct the density matrix, and therefore they have been
considered up to now \lq\lq {\it alternative complete}
descriptions of the dynamics \rq\rq \cite{barnettbook,mandel}.
However, the characteristic function can be considered an
alternative complete description of the dynamics if and only if it
is equivalent to the density matrix. I will prove in the following
that, contrarily to what has been believed until now, the
characteristic function and the density matrix descriptions of the
dynamics cannot be considered equivalent. To the best of the
author's knowledge this is the first time that the equivalence
between the characteristic function and the density matrix is
questioned, and an example pointing out the non-equivalence
between these two approaches is presented.

Let me begin by recalling the definition of the $p$-ordered
characteristic functions
\begin{eqnarray}
\chi (\xi, p)&=& {\rm Tr} \left[ \hat{\rho} \hat{D} (\xi)
\right]\exp\left( p |\xi|^2 /2 \right) \nonumber \\
&\equiv& {\rm Tr} \left[ \hat{\rho} \exp \left(\xi \hat{a}^{\dag}
- \xi^* \hat{a} \right) \right] \exp\left( p |\xi|^2 /2 \right),
\label{eq:chip}
\end{eqnarray}
where $\hat{D}(\xi)$ is the Glauber displacement operator and
$\hat{a}$ ($\hat{a}^{\dag}$) is the annihilation (creation)
operator of the quantum harmonic oscillator. In the previous
equation, the parameter $p$ assumes the values $p=1,0,-1$ in
correspondence to normal, symmetric and antinormal ordering of the
creation and annihilation operators. The two-dimensional Fourier
transform of $\chi(\xi,p)$ gives the Glauber-Sudarshan
$P$-representation for $p=1$, the Wigner function for $p=0$, and
the Husimi $Q$-function for $p=-1$ \cite{barnettbook}.

In what follows I will focus on the $p=0$ characteristic function
$\chi (\xi, p=0) \equiv \chi(\xi)$, known as simmetrically ordered
characteristic function (SCF) or quantum characteristic function.
Having in mind Eq. (\ref{eq:chip}) it is straightforward to prove
that the result obtained in this paper for the simmetrically
ordered characteristic function also applies to the other two
characteristic functions.


The SCF is always defined and it is, in general, a complex-valued
function satisfying the following properties:
\begin{equation}
\chi(\xi=0)=1; \hspace{1cm} \left| \chi(\xi) \right| \le 1.
\label{eq:cond}
\end{equation}
The first of the two properties is a consequence of the fact that
${\rm Tr}[\hat{\rho}]=1$, while the second stems from the fact
that $\chi(\xi)$ is the expectation value of the displacement
operator $\hat{D}(\xi)$ which is unitary, and therefore the
magnitude of its eigenvalues is unity. One of the advantages of
using the simmetrically ordered characteristic function is that
the analytic expression for the mean values of many observables of
interest may be calculated easily by means of the relation
\begin{eqnarray}
\langle a^{\dag m} a^n \rangle = \left. \left(\frac{d}{d
\xi}\right)^m \left(- \frac{d}{d \xi^*}\right)^n \chi(\xi)
\right|_{\xi=0}. \label{a}
\end{eqnarray}

In the literature the characteristic functions have been
extensively used to study the dynamics of both Markovian
\cite{barnettbook,mandel,Giovannetti04} and non-Markovian
\cite{IsarRev,Zurek89,Strunz03,PRAsolanalitica,Maniscalco04,Maniscalco04b}
open systems. It has also been shown that characteristic functions
may be used to establish observable conditions of nonclassicality
for the states of the quantized electromagnetic field
\cite{Vogel00}. In what follows I will show that there exist
situations in which the characteristic function description of an
open quantum system may lead to problems.

Let me consider the non-Markovian dynamics of a harmonic
oscillator interacting with a zero temperature reservoir. I
consider one of the most popular phenomenological model for this
systems, involving a memory kernel
\cite{petruccionebook,barnett,budini},
\begin{equation}
\frac{d \hat{\rho}(t)}{dt} = \int_0^t K(t-t') {\mathcal{L}}
\hat{\rho} (t') dt', \label{eq:mekernel}
\end{equation}
where $K(t-t')$ is the memory kernel and the Liouvillian operator
$\mathcal{L}$ is given by
\begin{equation}
{\mathcal{L}} \hat{\rho} = 2 \hat{a} \hat{\rho} \hat{a}^{\dag} -
\hat{a}^{\dag} \hat{a} \hat{\rho} - \hat{\rho} \hat{a}^{\dag}
\hat{a}.
\end{equation}
By applying the rules
\begin{eqnarray}
&& \hat{\rho}\rightarrow \chi (\xi), \hat{a}
\hat{\rho}\rightarrow\left(- \frac{d}{d \xi^*} -\frac{\xi}{2}
\right), \hat{a}^{\dag} \hat{\rho}\rightarrow\left( \frac{d}{d
\xi} -\frac{\xi^*}{2} \right), \nonumber
\\ &&  \hat{\rho}\hat{a} \rightarrow\left(-
\frac{d}{d \xi^*} +\frac{\xi}{2} \right), \hat{\rho}\hat{a}^{\dag}
\rightarrow\left( \frac{d}{d \xi} +\frac{\xi^*}{2} \right),
\end{eqnarray}
one may derive, from the master equation (\ref{eq:mekernel}),  the
corresponding integro-differential equation for $\chi(\xi)$
\begin{eqnarray}
\frac{\partial \chi(\xi, t)}{\partial t} &=& \int_0^t K(t-t')
\times \nonumber\\ &\times& \left[ - \left( \xi \frac{\partial
}{\partial \xi} + \xi^* \frac{\partial }{\partial \xi^*} \right) -
|\xi|^2 \right] \chi(\xi, t') dt'. \label{eq:chikernel}
\end{eqnarray}
Following a method developed in \cite{Sokolov02} for non-Markovian
Fokker-Plank equations, a formal solution of this
integro-differential equation may be obtained in form of an
integral decomposition involving the solution of the corresponding
Markovian problem, which is known in the literature (see, e.g.,
\cite{barnettbook}).

\begin{figure}
\includegraphics[width=7.5 cm,height=6.5 cm]{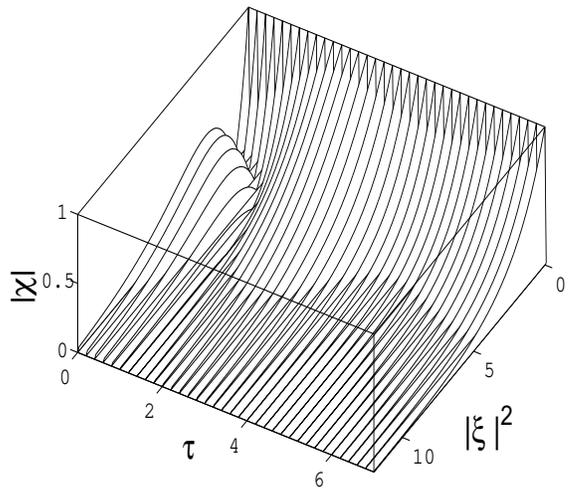}
\caption{Behavior of $\left|\chi(|\xi|^2 )\right|$, as given by
Eq. (\ref{eq:chisol}), at different times instants in the interval
$0<\tau<0$. The initial state is $\vert n =1 \rangle$ and
$g/\gamma =1$.} \label{fig:chi}
\end{figure}

Let me focus on the case of a memory kernel of exponential type
\begin{equation}
K(t-t')= g^2 e^{- \gamma |t-t'|}, \label{eq:kernel}
\end{equation}
with $g$ coupling strength and $\gamma$ decay constant of the
system-reservoir correlations. I consider, as initial state, a
Fock state $\vert n \rangle$, with $\vert n \rangle$ being the
eigenstates of the quantum number operator $\hat{n}=
\hat{a}^{\dag} \hat{a}$. The corresponding SCF reads as follows
\begin{equation}
\chi_n (\xi) = L_n(|\xi|^2) e^{-|\xi|^2/2},
\end{equation}
with $L_n(|\xi|^2)$ the Laguerre polynomial of order $n$. For
$n=1$, e.g., the initial simmetrically ordered characteristic
function is $\chi(\xi, 0) = (1-|\xi|^2) e^{-|\xi|^2/2}$, and it is
easy to verify by direct substitution  that
\begin{equation}
\chi(\xi,t)= \left[1- |\xi|^2  e^{-\gamma t/2} \left(\cos \Omega t
+ \frac{\gamma}{2 \Omega} \sin \Omega t \right) \right]
e^{-|\xi|^2/2}, \label{eq:chisol}
\end{equation}
is a solution of Eq. (\ref{eq:chikernel}), with $\Omega = \sqrt{2
g^2 -(\gamma/2)^2}$. Figure \ref{fig:chi} shows  the absolute
value of the simmetrically ordered characteristic function, as
given by Eq. (\ref{eq:chisol}), as a function of  $|\xi|^2$ at
different time instants and for $ g / \gamma=1$.
Looking at Eq. (\ref{eq:chisol}) it is easy to verify that, for
the parameters considered in the example,  the SCF is always
defined and it satisfies at all times and for each vale of $\xi$
the conditions given by Eqs. (\ref{eq:cond}), as one can also see
clearly from the figure.

Once the time evolution of the simmetrically ordered
characteristic function is known one can always reconstruct the
density matrix at all times, since $\hat{\rho}(t)$ can be obtained
from $\chi(\xi,t)$ by using the relation
\begin{equation}
\hat{\rho}(t)=\frac{1}{2 \pi} \int \chi(\xi,t) \hat{D}(\xi) d\xi
d\xi^*. \label{eq:rhochi}
\end{equation}
From this equation, keeping in mind that the diagonal elements of
the Glauber displacement operator are given by
\begin{equation}
\langle n \vert \hat{D}(\xi) \vert n \rangle = L_n(|\xi|^2)
e^{-|\xi|^2/2},
\end{equation}
and that $L_1(|\xi|^2) = 1-|\xi|^2$, one gets
\begin{eqnarray}
\rho_{11} (t) &\equiv& \langle n=1 \vert \hat{\rho} (t) \vert n=1
\rangle \nonumber \\
&=& e^{- \gamma t/2 } \left(\cos \Omega t + \frac{\gamma}{2
\Omega} \sin \Omega t \right). \label{eq:rho11}
\end{eqnarray}
The quantity $\rho_{11}(t)$ describes the decay of the population
of the initial state $\vert n=1 \rangle$ due to the interaction
with a zero-temperature reservoir. For $t \rightarrow \infty$,
$\hat{\rho}_{11} \rightarrow 0$ and $\hat{\rho}_{00} \rightarrow
1$, indicating that, due to dissipation, the state of the
oscillator passes from the initial excited Fock state $\vert n = 1
\rangle$ to the ground state. A close look to Eq.
(\ref{eq:rho11}), however, shows that the positivity of the
density matrix is clearly violated since $\hat{\rho}_{11}(t) < 0$
for some intervals of time. This situation has been discussed in
detail by Barnett and Stenholm in \cite{barnett}, where the risks
of an apparently physically well grounded memory kernel, as the
exponential one given by Eq. (\ref{eq:kernel}), were carefully
analyzed. The loss of  positivity shows up for strong or
intermediate coupling regime $ n g^2/\gamma \ge 1/8$
\cite{barnett} alerting us about the fact that the dynamics of the
system, for these values of the parameters, is unphysical.

It is worth stressing, however, that if one describes the time
evolution by using the simmetrically ordered characteristic
function, one does not realize that the equations of motion lose
physical sense because this function, contrarily to the density
matrix which violates one of its defining conditions (positivity),
continues to verify at all times the conditions given by Eqs.
(\ref{eq:cond}). Hence for non-Lindblad cases, when both the
complete positivity and the positivity conditions may be violated,
unphysical situations, such as the negativity of the density
matrix, may not show up in the dynamics of the SCF.

A careful analysis of Eq. (\ref{eq:chisol}) shows that only for
values of the ratio $g/\gamma$ such that, for certain intervals of
time, $\rho_{11}(t)< -1/2$ then the second of the conditions given
in Eqs. (\ref{eq:cond}) is violated for $|\xi|^2 \ll 1$. In
general, however, there is no correspondence between the loss of
the positivity condition, and therefore of complete positivity,
and the violation of one of the conditions defining the
simmetrically ordered characteristic function. Stated another way
the problem is the following. The fact that $\hat{D}(\xi)$ is an
unitary operator, together with the properties that $\hat{\rho}$
is a positive (then Hermitian) trace-class operator with trace
$1$, imply that $|\chi(\xi)|\le 1$, i.e. the second of Eq.
(\ref{eq:cond}). This is, however, only a necessary condition,
indeed I have shown in the paper that there exist situations in
correspondence of which the density matrix is not positive but
still $|\chi(\xi)|\le 1$. The crucial question is therefore, which
is the additional condition to be imposed on the SCF to ensure
that the operator $\hat{\rho}$, defined through Eq.
(\ref{eq:rhochi}), is a positive trace-class operator with trace
$1$? The answer to this question is very important since it would
allow to use safely the SCF in non-Lindblad cases. Until when this
condition is not found, one cannot claim that the density matrix
and the SCF are equivalent descriptions of the dynamics, as the
example given in this paper clearly indicates. The derivation of
the condition to be imposed on the simmetrically ordered
characteristic function to make it a useful tool in the
description of non-Lindblad type situations will be the object of
further study. However, presently, it seems to the author that
this question does not have a simple answer.

Let me conclude considering the behaviour of the Wigner function.
The Wigner function is the two-dimensional Fourier transform of
the simmetrically ordered characteristic function
\begin{equation}
W(\alpha)= \frac{1}{\pi^2} \int_{-\infty}^{\infty} d \xi d \xi^*
\chi(\xi) e^{\alpha \xi^*- \alpha^* \xi}. \label{eq:wig}
\end{equation}
The Wigner function is a real valued function satisfying the
condition $|W(\alpha)|\le 2/\pi$. Inserting Eq. (\ref{eq:chisol})
into Eq. (\ref{eq:wig}) we get
\begin{equation}
W(\alpha,t)=\frac{2}{\pi} e^{-2 |\alpha|^2} \left[ 1 + 2(2
|\alpha|^2 - 1) \rho_{11}(t) \right],
\end{equation}
with $\rho_{11}(t)$ given by Eq. (\ref{eq:rho11}). From the
previous equation one may verify that whenever $0 \le \rho_{11}(t)
\le 1$, then $|W(\alpha)|\le 2/\pi$, but the former inequality is
immediately violated in correspondence to a violation of the
positivity condition, in our example when $\rho_{11}(t)$ becomes
negative.

It is worth stressing the difference between the Wigner function,
which is the Fourier transform of the SCF, and the simmetrically
ordered characteristic function itself. While the first one
violates one of its defining conditions when the density matrix
loses positivity, the second one does not. In this sense it seems
that the simmetrically ordered characteristic function has
less\lq\lq physical meaning\rq\rq than the density matrix or the
Wigner function. It is worth noticing that, as a consequence of
the lack of a condition equivalent to the positivity of the
density matrix, each time one deals with non-Lindblad dynamics one
can use the SCF only if it is possible to derive the corresponding
density matrix, by means of Eq. (\ref{eq:rhochi}), and check its
positivity. In this paper I considered a rather easy example of
the dynamics for which both the density matrix and the SCF
solutions have simple analytic expressions. In general, however,
it is not obvious that, once the solution of the Fokker-Planck
equation for the SCF is known, one is also able to derive a useful
expression for the density matrix necessary to check the
positivity condition. This fact strongly limits the usability of
the SCF for the study of non-Lindblad dynamics. For this reason
new necessary and sufficient conditions establishing the
equivalence between the SCF and the density matrix are highly
desirable. Moreover, the answer to the open question posed in this
paper about the identification of a condition on the SCF
correspondent to positivity of the density matrix would shed light
on the physically meaningful ingredient allowing to consider the
description of an open quantum system via the SCF as a complete
description of its dynamics.

This work has been supported by the European Union's Transfer of
Knowledge project CAMEL (Grant No. MTKD-CT-2004-014427) and by the
Italian National Foundation Angelo Della Riccia. The author
gratefully acknowledges Antonino Messina, Francesco Petruccione,
Jyrki Piilo, and Stig Stenholm for the useful comments and
suggestions about the paper.

\end{document}